\documentclass[10pt]{article}
\usepackage{graphicx}
\usepackage{amsmath}
\usepackage{amssymb}
\usepackage{caption2}
\begin{document}
\begin{center}
{\large {\bf \sc{  The strange cousin of the $Z_c(4020/4025)$ as tetraquark state  }}} \\[2mm]
Zhi-Gang  Wang \footnote{E-mail: zgwang@aliyun.com.  }     \\
 Department of Physics, North China Electric Power University, Baoding 071003, P. R. China
\end{center}

\begin{abstract}
Motivated  by the analogous properties of the $Z_c(3900/3885)$ and $Z_{cs}(3985/4000)$, we tentatively assign the $Z_c(4020/4025)$ as the $A\bar{A}$-type hidden-charm tetraquark state with the $J^{PC}=1^{+-}$, where the $A$ denotes the axialvector diquark states, and explore the $A\bar{A}$-type tetraquark states without strange, with strange and with hidden-strange via the QCD sum rules in a consistent way. Then we   explore the hadronic  coupling constants in the two-body strong decays  of the tetraquark states without strange and with strange via the QCD sum rules based on rigorous quark-hadron duality, and acquire the partial decay widths and total decay widths.  The present calculations  support assigning the $Z_c(4020/4025)$ as  the  $A\bar{A}$-type tetraquark state with the $J^{PC}=1^{+-}$,  while the predictions for its strange cousin $Z_{cs}$ state can be confronted to the experimental data in the future.
 \end{abstract}



\section{Introduction}

 In 2013, the BESIII collaboration  observed
the $Z^{\pm}_c(4025)$  in the $\pi^\mp$ recoil mass spectrum  in the process $e^+e^- \to (D^{*} \bar{D}^{*})^{\pm} \pi^\mp$,  the measured   Breit-Wigner mass and width are $M=(4026.3\pm2.6\pm3.7)\,\rm{MeV}$  and $\Gamma=(24.8\pm5.6\pm7.7)\,\rm{MeV}$, respectively \cite{BES1308}. Two years later, the BESIII collaboration  observed its neutral partner  $Z^{0}_c(4025)$  in the $\pi^0$ recoil mass spectrum  in the process $e^+e^- \to (D^{*} \bar{D}^{*})^{0} \pi^0$,   the  measured  Breit-Wigner mass and width are  $M=(4025.5^{+2.0}_{-4.7}\pm3.1)\,\rm{MeV}$  and $\Gamma=(23.0\pm 6.0\pm 1.0)\,\rm{MeV}$, respectively \cite{BES1507}. The masses and widths of the charged structures  $Z_c^{\pm}(4025)$ and neutral structure $Z_c^0(4025)$ are consistent with each other.
Also in 2013, the  BESIII collaboration observed the  $Z_c^\pm(4020)$   in the $\pi^\pm h_c$ mass spectrum in the process $e^+e^- \to \pi^+\pi^- h_c$,  the measured   Breit-Wigner   mass and width are  $M=(4022.9\pm 0.8\pm 2.7)\,\rm{MeV}$   and $\Gamma=(7.9\pm 2.7\pm 2.6)\,\rm{MeV}$, respectively  \cite{BES1309}.  The $Z_c(4020)$ and $Z_c(4025)$ are assigned to be the same particle by the Particle Data Group, and listed in the Review of Particle Physics  as $X(4020)$ \cite{PDG}, although the widths differ from each other considerably.

The spin and parity have not been measured yet, the S-wave $D^{*} \bar{D}^{*}$ systems  have the quantum numbers $J^{PC}=0^{++}$, $1^{+-}$, $2^{++}$,  the S-wave $ \pi^\pm h_c$ systems have the quantum numbers $J^{PC}=1^{--}$,  the P-wave $\pi^{\pm} h_c$ systems  have the quantum numbers $J^{PC}=0^{++}$, $1^{+-}$, $2^{++}$, we can tentatively assign the quantum numbers $J^{PC}=1^{+-}$ for the $Z_c(4020/4025)$. According to the nearby $D^*\bar{D}^*$ threshold, one maybe  expect to assign the $Z_c(4020/4025)$ as the
tetraquark molecular state \cite{Molecule-1,Molecule-2,Molecule-3,Molecule-4,Molecule-5,Molecule-6,Molecule-7,ZhangJR3900}. In the picture of tetraquark states, the $Z_c(4020/4025)$ can be assigned as the $A\bar{A}$-type tetraquark state with the $J^{PC}=1^{+-}$ \cite{Tetraquark-Qiao,WangTetraquarkCTP,Zc4025-EPJC-Mc}, while the $Z_c(3900)$ can be assigned to be the $S\bar{A}-A\bar{S}$ type tetraquark state
according to the calculations via the QCD sum rules \cite{WangHuang-PRD}, where the $S$ and $A$ represent the scalar and axialvector diquark states, respectively.

In 2020, the BESIII collaboration observed a structure $Z_{cs}^-(3985)$ in the $K^{+}$ recoil-mass spectrum with the significance of  5.3 $\sigma$ in the processes of the $e^+e^-\to K^+ (D_s^- D^{*0} + D^{*-}_s D^0)$ \cite{BES3985}.   The measured Breit-Wigner  mass and width  are  $M=3985.2^{+2.1}_{-2.0}\pm1.7\,\rm{MeV}$   and $\Gamma=13.8^{+8.1}_{-5.2}\pm4.9\,\rm{MeV}$, respectively  \cite{BES3985}.
In 2021, the LHCb collaboration observed two new  exotic states $Z_{cs}^+(4000)$ and $Z_{cs}^+(4220)$ in the $J/\psi K^+$  mass spectrum  in the process  $B^+ \to J/\psi \phi K^+$ \cite{LHCb-Zcs4000}.  The most significant state, $Z_{cs}^+(4000)$, has the Breit-Wigner  mass and width $M=4003 \pm 6 {}^{+4}_{-14}\,\rm{MeV}$ and $\Gamma=131 \pm 15 \pm 26\,\rm{MeV}$, respectively,  and the spin-parity $J^P =1^+$ \cite{LHCb-Zcs4000}. Although in both the pictures of tetraquark states and molecular states, we can reproduce the mass of the $Z_{cs}(3985/4000)$ with the QCD sum rules \cite{Zcs-CFQiao,WZG-Zcs3985-tetra,Zcs3985-Azizi,Ozdem-Zcs-tetra-mole,WZG-Zcs3985-mole,Lee:2008uy,Dias:2013qga,ChenHX-Zcs3985}, direct calculations of the decay widths based on the QCD sum rules support assigning the $Z_{cs}(3985)$ and $Z_{cs}(4000)$ to be the hidden-charm tetraquark state and molecular state with the $J^{PC}=1^{+-}$, respectively, or at least, the $Z_{cs}(3985)$ maybe have large diquark-antidiquark type Fock component, while the $Z_{cs}(4000)$ maybe have large color-singlet-color-singlet type Fock component \cite{WZG-Zcs3985-decay}.

The $Z_c(3900/3885)$ and $Z_{cs}(3985/4000)$ are cousins and  have analogous decay modes,
\begin{eqnarray}
Z^{\pm}_c(3900) &\to & J/\psi \pi^\pm \, , \nonumber \\
Z^{+}_{cs}(4000) &\to & J/\psi K^+ \, ,
\end{eqnarray}
\begin{eqnarray}
Z_c^{\pm}(3885)&\to & (D\bar{D}^*)^\pm \, , \nonumber \\
Z_{cs}^{-}(3985) &\to & D_s^- D^{*0}\, , \,  D^{*-}_s D^0 \, ,
\end{eqnarray}
we expect that the $Z_{c}(4020/4025)$ also have strange cousins $Z_{cs}$, and they have analogous decay modes, the $Z_{cs}$ states  may be observed in the decays to the final states $D^*\bar{D}_s^*$, $D_s^*\bar{D}^*$, $h_cK$, etc. In this work, we tentatively assign the $Z_c(4020/4025)$ as the $A\bar{A}$-type hidden-charm tetraquark state with the $J^{PC}=1^{+-}$,  and  extend our previous work to study the mass and width of its strange cousin with the QCD sum rules \cite{WZG-Zcs3985-tetra,WZG-Zcs3985-mole,WZG-Zcs3985-decay,WZG-tetra-spectrum}, the predictions can be confronted to the experimental data in the future, and make contribution in disentangling   the pictures of tetraquark states and molecular states.
 As a byproduct, we obtain the mass of the hidden-strange/charm  tetraquark state and the partial decay widths of the $Z_c(4020/4025)$.

The article is arranged as follows:  we derive the QCD sum rules for the masses and pole residues of  the  $A\bar{A}$-type tetraquark states without strange, with strange, and with hidden-strange in section 2; in section 3, we derive the QCD sum rules for the hadronic coupling constants in the decays of   the  $Z_c$ and $Z_{cs}$ states; section 4 is reserved for our conclusion.

\section{QCD sum rules for  the  $Z_{c}$, $Z_{cs}$, $Z_{cs\bar{s}}$ tetraquark states with the $J^{PC}=1^{+-}$}
Firstly, we write down  the two-point correlation functions  $\Pi_{\mu\nu\alpha\beta}(p)$  in the QCD sum rules,
\begin{eqnarray}
\Pi_{\mu\nu\alpha\beta}(p)&=&i\int d^4x e^{ip \cdot x} \langle0|T\left\{J_{\mu\nu}(x)J_{\alpha\beta}^{\dagger}(0)\right\}|0\rangle \, ,
\end{eqnarray}
where $J_{\mu\nu}(x)=J_{\mu\nu}^{u\bar{d}}(x)$, $J_{\mu\nu}^{u\bar{s}}(x)$, $J_{\mu\nu}^{s\bar{s}}(x)$,
\begin{eqnarray}
J^{u\bar{d}}_{\mu\nu}(x)&=&\frac{\varepsilon^{ijk}\varepsilon^{imn}}{\sqrt{2}}\Big\{u^T_j(x) C\gamma_\mu c_k(x) \bar{d}_m(x) \gamma_\nu C \bar{c}^T_n(x)  -u^T_j(x) C\gamma_\nu c_k(x) \bar{d}_m(x) \gamma_\mu C \bar{c}^T_n(x) \Big\} \, , \nonumber \\
J^{u\bar{s}}_{\mu\nu}(x)&=&\frac{\varepsilon^{ijk}\varepsilon^{imn}}{\sqrt{2}}\Big\{u^T_j(x) C\gamma_\mu c_k(x) \bar{s}_m(x) \gamma_\nu C \bar{c}^T_n(x)  -u^T_j(x) C\gamma_\nu c_k(x) \bar{s}_m(x) \gamma_\mu C \bar{c}^T_n(x) \Big\} \, , \nonumber \\
J^{s\bar{s}}_{\mu\nu}(x)&=&\frac{\varepsilon^{ijk}\varepsilon^{imn}}{\sqrt{2}}\Big\{s^T_j(x) C\gamma_\mu c_k(x) \bar{s}_m(x) \gamma_\nu C \bar{c}^T_n(x)  -s^T_j(x) C\gamma_\nu c_k(x) \bar{s}_m(x) \gamma_\mu C \bar{c}^T_n(x) \Big\} \, ,
 \end{eqnarray}
where the $i$, $j$, $k$, $m$, $n$ are color indexes, the $C$ is the charge conjugation matrix \cite{Zc4025-EPJC-Mc,WZG-tetra-spectrum}. We choose the currents $J_{\mu\nu}^{u\bar{d}}(x)$, $J_{\mu\nu}^{u\bar{s}}(x)$ and $J_{\mu\nu}^{s\bar{s}}(x)$ to explore the hidden-charm tetraquark states without strange, with strange, and with hidden-strange, respectively.

At the hadronic side, we  isolate the ground state contributions of the  hidden-charm tetraquark states with the $J^{PC}=1^{+-}$ and $1^{--}$ explicitly, and acquire  the results,
\begin{eqnarray}
\Pi_{\mu\nu\alpha\beta}(p)&=&\frac{\lambda_{ Z}^2}{M_{Z}^2-p^2}\left(p^2g_{\mu\alpha}g_{\nu\beta} -p^2g_{\mu\beta}g_{\nu\alpha} -g_{\mu\alpha}p_{\nu}p_{\beta}-g_{\nu\beta}p_{\mu}p_{\alpha}+g_{\mu\beta}p_{\nu}p_{\alpha}+g_{\nu\alpha}p_{\mu}p_{\beta}\right) \nonumber\\
&&+\frac{\lambda_{ Y}^2}{M_{Y}^2-p^2}\left( -g_{\mu\alpha}p_{\nu}p_{\beta}-g_{\nu\beta}p_{\mu}p_{\alpha}+g_{\mu\beta}p_{\nu}p_{\alpha}+g_{\nu\alpha}p_{\mu}p_{\beta}\right) +\cdots \, \, ,
\end{eqnarray}
where the $Z$ and $Y$ denote  the  tetraquark states with the $J^{PC}=1^{+-}$ and $1^{--}$, respectively,  the  pole residues  $\lambda_{Z}$ and $\lambda_{Y}$ are defined by
\begin{eqnarray}
  \langle 0|\eta_{\mu\nu}(0)|Z(p)\rangle &=& \lambda_{Z} \, \varepsilon_{\mu\nu\alpha\beta} \, \zeta^{\alpha}p^{\beta}\, , \nonumber\\
 \langle 0|\eta_{\mu\nu}(0)|Y(p)\rangle &=& \lambda_{Y} \left(\zeta_{\mu}p_{\nu}-\zeta_{\nu}p_{\mu} \right)\, ,
\end{eqnarray}
the  $\zeta_\mu$ are the polarization vectors of the tetraquark states.
We can rewrite the correlation functions  $\Pi_{\mu\nu\alpha\beta}(p)$ into the form,
\begin{eqnarray}
\Pi_{\mu\nu\alpha\beta}(p)&=&\Pi_Z(p^2)\left(p^2g_{\mu\alpha}g_{\nu\beta} -p^2g_{\mu\beta}g_{\nu\alpha} -g_{\mu\alpha}p_{\nu}p_{\beta}-g_{\nu\beta}p_{\mu}p_{\alpha}+g_{\mu\beta}p_{\nu}p_{\alpha}+g_{\nu\alpha}p_{\mu}p_{\beta}\right) \nonumber\\
&&+\Pi_Y(p^2)\left( -g_{\mu\alpha}p_{\nu}p_{\beta}-g_{\nu\beta}p_{\mu}p_{\alpha}+g_{\mu\beta}p_{\nu}p_{\alpha}+g_{\nu\alpha}p_{\mu}p_{\beta}\right) \, ,
\end{eqnarray}
according to Lorentz covariance.

We project out the components $\Pi_Z(p^2)$ and $\Pi_Y(p^2)$ by  the tensors $P_{A,p}^{\mu\nu\alpha\beta}$ and $P_{V,p}^{\mu\nu\alpha\beta}$,
\begin{eqnarray}
\widetilde{\Pi}_Z(p^2)&=&p^2\Pi_Z(p^2)=P_{A,p}^{\mu\nu\alpha\beta}\Pi_{\mu\nu\alpha\beta}(p) \, , \nonumber\\
\widetilde{\Pi}_Y(p^2)&=&p^2\Pi_Y(p^2)=P_{V,p}^{\mu\nu\alpha\beta}\Pi_{\mu\nu\alpha\beta}(p) \, ,
\end{eqnarray}
where
\begin{eqnarray}
P_{A,p}^{\mu\nu\alpha\beta}&=&\frac{1}{6}\left( g^{\mu\alpha}-\frac{p^\mu p^\alpha}{p^2}\right)\left( g^{\nu\beta}-\frac{p^\nu p^\beta}{p^2}\right)\, , \nonumber\\
P_{V,p}^{\mu\nu\alpha\beta}&=&\frac{1}{6}\left( g^{\mu\alpha}-\frac{p^\mu p^\alpha}{p^2}\right)\left( g^{\nu\beta}-\frac{p^\nu p^\beta}{p^2}\right)-\frac{1}{6}g^{\mu\alpha}g^{\nu\beta}\, .
\end{eqnarray}

 We accomplish  the operator product expansion  up to the vacuum condensates of dimension 10, and take account of the vacuum condensates $\langle \bar{q}q\rangle$, $\langle \frac{\alpha_s GG}{\pi}\rangle$,
$\langle \bar{q}g_s\sigma Gq\rangle$, $\langle \bar{q}q\rangle^2$,
$ \langle \bar{q}q\rangle \langle \frac{\alpha_s GG}{\pi}\rangle$, $\langle\bar{q}q\rangle\langle \bar{q}g_s\sigma Gq\rangle$,
$\langle \bar{q}g_s\sigma Gq\rangle^2$, $\langle \bar{q}q\rangle^2 \langle \frac{\alpha_s GG}{\pi}\rangle$, where $q=u$, $d$ or $s$, just like in previous works \cite{WangTetraquarkCTP,Zc4025-EPJC-Mc,WangHuang-PRD,WZG-Zcs3985-tetra,WZG-Zcs3985-mole},
 and project out the components,
  \begin{eqnarray}
\widetilde{\Pi}_Z(p^2)&=&P_{A,p}^{\mu\nu\alpha\beta}\Pi_{\mu\nu\alpha\beta}(p) \, , \nonumber\\
\widetilde{\Pi}_Y(p^2)&=&P_{V,p}^{\mu\nu\alpha\beta}\Pi_{\mu\nu\alpha\beta}(p) \, ,
\end{eqnarray}
at the QCD side. In the present work, we only interest in the component $\widetilde{\Pi}_Z(p^2)$, as we investigate the axialvector tetraquark states.  We take
the truncations $n\leq 10$ and $k\leq 1$ in a consistent way,
the operators of the orders $\mathcal{O}( \alpha_s^{k})$ with $k> 1$ are  discarded.   The operators in the condensates $\langle g_s^3 GGG\rangle$, $\langle \frac{\alpha_s GG}{\pi}\rangle^2$,
 $\langle \frac{\alpha_s GG}{\pi}\rangle\langle \bar{q} g_s \sigma Gq\rangle$ are of the orders $\mathcal{O}( \alpha_s^{3/2})$, $\mathcal{O}(\alpha^2_s)$, $\mathcal{O}( \alpha_s^{3/2})$, respectively, and play tiny roles, and can be ignored  safely \cite{ZhangJR3900,WangXW-afs-afs}.

We obtain the QCD spectral densities $\rho_Z(s)$ through dispersion relation,
\begin{eqnarray}
\rho_Z(s)&=&\frac{{\rm Im}\widetilde{\Pi}_Z(s)}{\pi}\, ,
\end{eqnarray}
 suppose  the quark-hadron duality below the continuum thresholds  $s_0$ and accomplish Borel transform  in regard  to
the variable $P^2=-p^2$ to obtain  the QCD sum rules:
\begin{eqnarray} \label{QCDSR-two}
\tilde{\lambda}^2_{Z}\, \exp\left(-\frac{M^2_Z}{T^2}\right)= \int_{4m_c^2}^{s_0} ds\, \rho_Z(s) \, \exp\left(-\frac{s}{T^2}\right) \, ,
\end{eqnarray}
where $\tilde{\lambda}_Z=\lambda_{Z}M_{Z}$.

We differentiate   Eq.\eqref{QCDSR-two}  in regard to  $\frac{1}{T^2}$, eliminate the re-defined
 pole residues $\tilde{\lambda}_{Z}$, and obtain the QCD sum rules for
 the masses of those axialvector hidden-charm  tetraquark states,
 \begin{eqnarray}
 M^2_{Z}&=& \frac{\int_{4m_c^2}^{s_0} ds\,\frac{d}{d \left(-1/T^2\right)}\,\rho_Z(s)\,\exp\left(-\frac{s}{T^2}\right)}{\int_{4m_c^2}^{s_0} ds\, \rho_Z(s)\,\exp\left(-\frac{s}{T^2}\right)}\,   .
\end{eqnarray}

We take  the standard values of the vacuum condensates,
$\langle
\bar{q}q \rangle=-(0.24\pm 0.01\, \rm{GeV})^3$, $\langle\bar{s}s\rangle=(0.8\pm0.1)\langle\bar{q}q\rangle$,
$\langle\bar{q}g_s\sigma G q \rangle=m_0^2\langle \bar{q}q \rangle$,
$\langle\bar{s}g_s\sigma G s \rangle=m_0^2\langle \bar{s}s \rangle$,
$m_0^2=(0.8 \pm 0.1)\,\rm{GeV}^2$     at the   energy scale  $\mu=1\, \rm{GeV}$
\cite{SVZ79,Reinders85,Colangelo-Review},  and take the $\overline{MS}$ quark masses $m_{c}(m_c)=(1.275\pm0.025)\,\rm{GeV}$ and $m_s(\mu=2\,\rm{GeV})=(0.095\pm0.005)\,\rm{GeV}$ from the Particle Data Group \cite{PDG}. We set $m_q=m_u=m_d=0$ and take account of
the energy-scale dependence of  the input parameters,
\begin{eqnarray}
\langle\bar{q}q \rangle(\mu)&=&\langle\bar{q}q \rangle({\rm 1GeV})\left[\frac{\alpha_{s}({\rm 1GeV})}{\alpha_{s}(\mu)}\right]^{\frac{12}{33-2n_f}}\, , \nonumber\\
\langle\bar{s}s \rangle(\mu)&=&\langle\bar{s}s \rangle({\rm 1GeV})\left[\frac{\alpha_{s}({\rm 1GeV})}{\alpha_{s}(\mu)}\right]^{\frac{12}{33-2n_f}}\, , \nonumber\\
 \langle\bar{q}g_s \sigma Gq \rangle(\mu)&=&\langle\bar{q}g_s \sigma Gq \rangle({\rm 1GeV})\left[\frac{\alpha_{s}({\rm 1GeV})}{\alpha_{s}(\mu)}\right]^{\frac{2}{33-2n_f}}\, , \nonumber\\
  \langle\bar{s}g_s \sigma Gs \rangle(\mu)&=&\langle\bar{s}g_s \sigma Gs \rangle({\rm 1GeV})\left[\frac{\alpha_{s}({\rm 1GeV})}{\alpha_{s}(\mu)}\right]^{\frac{2}{33-2n_f}}\, , \nonumber\\
 m_c(\mu)&=&m_c(m_c)\left[\frac{\alpha_{s}(\mu)}{\alpha_{s}(m_c)}\right]^{\frac{12}{33-2n_f}} \, ,\nonumber\\
 m_s(\mu)&=&m_s({\rm 2GeV})\left[\frac{\alpha_{s}(\mu)}{\alpha_{s}({\rm 2GeV})}\right]^{\frac{12}{33-2n_f}} \, ,\nonumber\\
\alpha_s(\mu)&=&\frac{1}{b_0t}\left[1-\frac{b_1}{b_0^2}\frac{\log t}{t} +\frac{b_1^2(\log^2{t}-\log{t}-1)+b_0b_2}{b_0^4t^2}\right]\, ,
\end{eqnarray}
 from the renormalization group equation, where   $t=\log \frac{\mu^2}{\Lambda_{QCD}^2}$, $b_0=\frac{33-2n_f}{12\pi}$, $b_1=\frac{153-19n_f}{24\pi^2}$, $b_2=\frac{2857-\frac{5033}{9}n_f+\frac{325}{27}n_f^2}{128\pi^3}$,  $\Lambda_{QCD}=210\,\rm{MeV}$, $292\,\rm{MeV}$  and  $332\,\rm{MeV}$ for the flavors  $n_f=5$, $4$ and $3$, respectively  \cite{PDG,Narison-mix}, we choose the flavor numbers $n_f=4$ as there are $u$, $d$, $s$ and $c$ quarks.

Just as in our previous works,  we acquire  the acceptable energy scales of the QCD spectral densities  for the hidden-charm  tetraquark states according to the energy scale formula,
\begin{eqnarray}
\mu &=&\sqrt{M^2_{X/Y/Z}-(2{\mathbb{M}}_c)^2} \, ,
 \end{eqnarray}
with the effective $c$-quark mass  ${\mathbb{M}}_c=1.82\,\rm{GeV}$ \cite{Zc4025-EPJC-Mc,WZG-fomula-mole-1,WZG-formula-mole-2,WZG-fromula-tetra}. Furthermore, we take account of the $SU(3)$ mass-breaking effects according to modified energy scale formula,
\begin{eqnarray}
\mu &=&\sqrt{M^2_{X/Y/Z}-(2{\mathbb{M}}_c)^2}-k\,{\mathbb{M}}_s \, ,
 \end{eqnarray}
 where the ${\mathbb{M}}_s$ is the effective $s$-quark mass and fitted to be $0.2\,\rm{GeV}$  \cite{WZG-Zcs3985-mole}, the $k$ is the number of the valence $s$-quarks.

\begin{figure}
\centering
\includegraphics[totalheight=6cm,width=7cm]{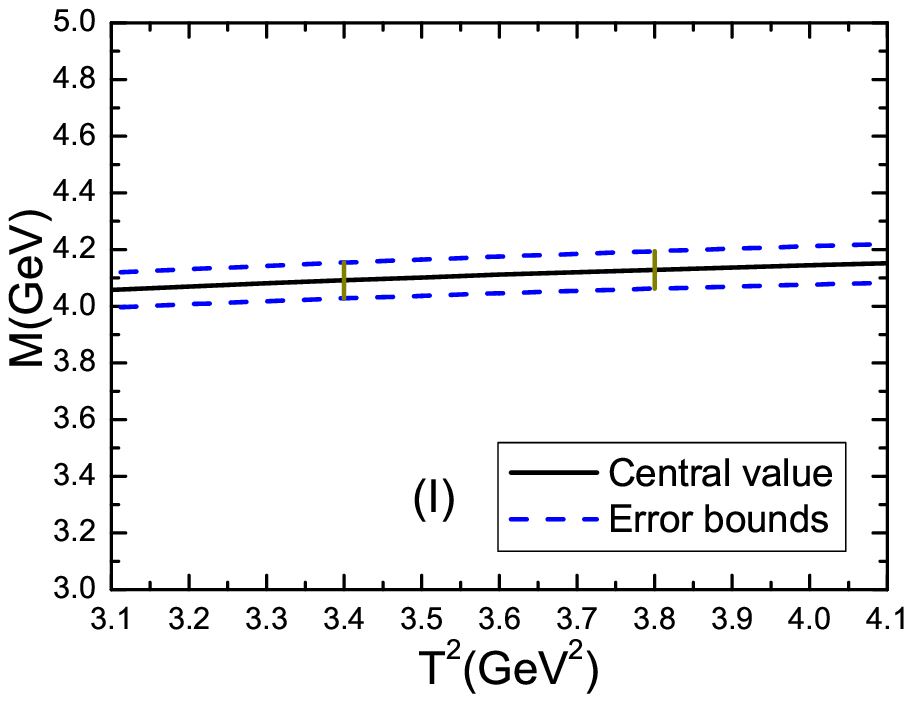}
\includegraphics[totalheight=6cm,width=7cm]{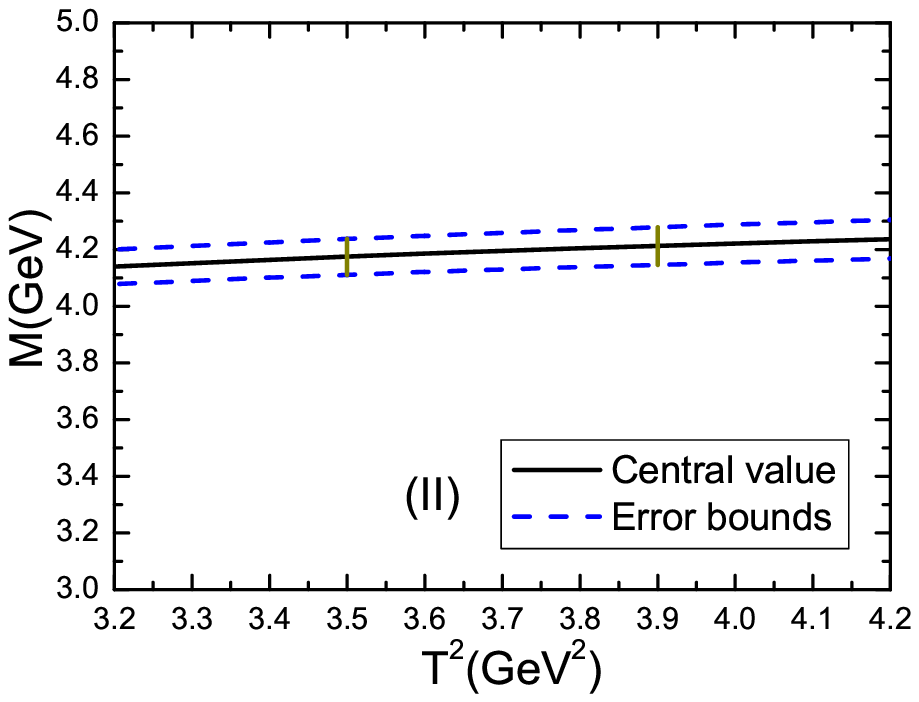}
  \caption{ The masses of the tetraquark  states with variations of the Borel parameters $T^2$, where the (I) and (II) correspond  to the $Z_{cs}$ and $Z_{cs\bar{s}}$, respectively, the regions between the two vertical lines are the Borel windows.    }\label{mass-Zc-Zcs}
\end{figure}

We search for the suitable Borel parameters $T^2$ and continuum threshold parameters $s_0$ to satisfy the two criteria (pole or ground state dominance and convergence of the operator product expansion) via trial and error.   The  Borel parameters, continuum threshold parameters, energy scales of the QCD spectral densities, pole contributions  and contributions of the vacuum condensates of dimension 10  are shown in Table \ref{Borel-mass}. From the Table, we can see plainly that the modified energy scale formula is satisfied very well. Then we take account of the uncertainties of the input parameters, and acquire the masses and pole residues of the hidden-charm tetraquark states without strange, with strange and with hidden-strange having  the quantum numbers $J^{PC}=1^{+-}$, which are also shown in Table \ref{Borel-mass}. In Fig.\ref{mass-Zc-Zcs}, we plot the masses of the $Z_{cs}$ and $Z_{cs\bar{s}}$ with variations of the Borel parameters, from the figure, we can see that there appear platforms in the Borel windows indeed, it is reliable to extract the tetraquark masses.

\begin{table}
\begin{center}
\begin{tabular}{|c|c|c|c|c|c|c|c|c|}\hline\hline
                &$T^2 (\rm{GeV}^2)$ &$\sqrt{s_0}(\rm GeV) $ &$\mu(\rm{GeV})$ &pole        &$|D(10)|$ &$M_Z(\rm{GeV})$  &$\tilde{\lambda}_Z(10^{-2}\rm{GeV}^5)$  \\ \hline

$Z_c$           &$3.3-3.7$          &$4.6\pm0.1$            &$1.7$           &$(40-59)\%$ &$\ll 1\%$ & $4.02\pm0.09$   &$3.00\pm0.45 $  \\ \hline

$Z_{cs}$        &$3.4-3.8$          &$4.7\pm0.1$            &$1.7$           &$(41-60)\%$ &$\ll 1\%$ & $4.11\pm0.08$   &$3.49\pm0.51$  \\ \hline

$Z_{cs\bar{s}}$ &$3.5-3.9$          &$4.8\pm0.1$            &$1.7$           &$(42-61)\%$ &$\ll 1\%$ & $4.20\pm0.09$   &$4.00\pm0.58$  \\ \hline
\end{tabular}
\end{center}
\caption{ The Borel parameters, continuum threshold parameters, energy scales, pole contributions,  contributions of the vacuum condensates of dimension $10$, masses and pole residues  for the axialvector tetraquark states. }\label{Borel-mass}
\end{table}

The present prediction  $ M_{Z_c}=\left(4.02\pm0.09\right) \, \rm{GeV}$ (also in Ref.\cite{WZG-tetra-spectrum}) is consistent with the experimental values $M_{Z_c^\pm}=(4026.3\pm2.6\pm3.7)\,\rm{MeV}$, $M_{Z_c^\pm}=(4022.9\pm 0.8\pm 2.7)\,\rm{MeV}$, $M_{Z_c^0}=(4025.5^{+2.0}_{-4.7}\pm3.1)\,\rm{MeV}$
from the BESIII collaboration \cite{BES1308,BES1507,BES1309},    which supports assigning the $Z_c(4020/4025)$   to be  the $J^{PC}=1^{+-}$  $A\bar{A}$-type tetraquark state. We cannot assign a hadron unambiguously with the mass alone, we have to calculate the partial decay widths and total width to make more robust assignment.

\section{Decay widths of the  $Z_c$ and $Z_{cs}$ states with the QCD sum rules}

We  investigate the two-body strong decays $Z_{cs}\to h_cK$, $J/\psi K$, $\eta_cK^{*}$ with  the three-point correlation functions $\Pi_{\alpha\beta\mu\nu}(p,q)$, $\Pi_{\alpha\mu\nu}^{1}(p,q)$ and $\Pi_{\alpha\mu\nu}^{2}(p,q)$, respectively,
\begin{eqnarray}
\Pi_{\alpha\beta\mu\nu}(p,q)&=&i^2\int d^4xd^4y \, e^{ipx}e^{iqy}\, \langle 0|T\left\{J_{\alpha\beta}^{h_c}(x)J_5^{K}(y)J^{u\bar{s}\dagger}_{\mu\nu}(0)\right\}|0\rangle\, ,  \nonumber \\
\Pi_{\alpha\mu\nu}^{1}(p,q)&=&i^2\int d^4xd^4y \, e^{ipx}e^{iqy}\, \langle 0|T\left\{J_\alpha^{J/\psi}(x)J_5^{K}(y)J^{u\bar{s}\dagger}_{\mu\nu}(0)\right\}|0\rangle \, , \nonumber \\
\Pi_{\alpha\mu\nu}^{2}(p,q)&=&i^2\int d^4xd^4y\, e^{ipx}e^{iqy}\, \langle 0|T\left\{J_5^{\eta_c}(x)J_\alpha^{K^*}(y)J^{u\bar{s}\dagger}_{\mu\nu}(0)\right\}|0\rangle \, ,
\end{eqnarray}
where the currents
\begin{eqnarray}
J_{\alpha\beta}^{h_c}(x)&=&\bar{c}(x)\sigma_{\alpha\beta} c(x) \, ,\nonumber \\
J_\mu^{J/\psi}(x)&=&\bar{c}(x)\gamma_\mu c(x) \, ,\nonumber \\
J_5^{K}(y)&=&\bar{u}(y)i\gamma_5 s(y) \, , \nonumber \\
J_5^{\eta_c}(x)&=&\bar{c}(x)i\gamma_5 c(x) \, ,\nonumber \\
J_\mu^{K^*}(y)&=&\bar{u}(y)\gamma_\mu s(y) \, ,
\end{eqnarray}
interpolate the mesons $h_c$, $J/\psi$, $K$, $\eta_c$ and $K^*$, respectively, with the simple replacement $s\to d$, we obtain the corresponding ones for the
$Z_c$ tetraquark state.

We insert  a complete set of intermediate hadronic states having possible (non-vanishing) couplings  with the current operators into the three-point correlation functions, and  isolate the ground state contributions explicitly,
\begin{eqnarray}\label{Hadron-CT-1}
\Pi^{\alpha\beta\mu\nu}
(p,q)&=& \lambda_K f_{h}\varepsilon^{\alpha\beta\alpha^\prime\beta^\prime} \xi_{\alpha^\prime}  p_{\beta^\prime} \lambda_{Z}
\varepsilon^{\mu\nu\mu^\prime\nu^\prime}\zeta^{*}_{\mu^\prime}p^{\prime}_{\nu^\prime} \frac{-iG_{Zh K}\varepsilon^{\rho\sigma\lambda\tau}p_\rho\xi^{*}_\sigma p^{\prime}_\lambda\zeta_\tau}{(M_{Z}^2-p^{\prime2})(M_{h}^2-p^2)(M_{K}^2-q^2)}  \nonumber\\
&&+ \lambda_K f_{h}\varepsilon^{\alpha\beta\alpha^\prime\beta^\prime}\xi_{\alpha^\prime}p_{\beta^\prime}\lambda_{Y}
\left(\zeta^{*\mu }p^{\prime\nu }-\zeta^{*\nu }p^{\prime\mu } \right)\frac{-G_{Y h K}\xi^{*} \cdot \zeta }{(M_{Y}^2-p^{\prime2})(M_{h}^2-p^2)(M_{K}^2-q^2)}  \nonumber\\
&&+\lambda_K f_{J/\psi}^T\left(\xi^{\alpha}p^{\beta}-\xi^{\beta}p^{\alpha}\right)\lambda_{Z}
\varepsilon^{\mu\nu\mu^\prime\nu^\prime}\zeta^{*}_{\mu^\prime}p^{\prime}_{\nu^\prime} \frac{-G_{ZJ/\psi K}\xi^{*} \cdot \zeta}{(M_{Z}^2-p^{\prime2})(M_{J/\psi}^2-p^2)(M_{K}^2-q^2)}  \nonumber\\
&&+\lambda_K f_{J/\psi}^T\left(\xi^{\alpha}p^{\beta}-\xi^{\beta}p^{\alpha}\right)\lambda_{Y}
\left(\zeta^{*\mu }p^{\prime\nu }-\zeta^{*\nu }p^{\prime\mu } \right)\frac{-iG_{YJ/\psi K}\varepsilon^{\rho\sigma\lambda\tau}p_\rho\xi^{*}_\sigma p^{\prime}_\lambda\zeta_\tau}{(M_{Y}^2-p^{\prime2})(M_{J/\psi}^2-p^2)(M_{K}^2-q^2)}  \nonumber\\
&&+ \cdots\, ,
\end{eqnarray}

\begin{eqnarray}\label{Hadron-CT-2}
\Pi^{\alpha\mu\nu}_1(p,q)&=& \lambda_K \lambda_{J/\psi}\xi^{\alpha } \lambda_{Z}
\varepsilon^{\mu\nu\mu^\prime\nu^\prime}\zeta^{*}_{\mu^\prime}p^{\prime}_{\nu^\prime} \frac{-G_{ZJ/\psi K}\xi^{*} \cdot  \zeta }{(M_{Z}^2-p^{\prime2})(M_{J/\psi}^2-p^2)(M_{K}^2-q^2)}  \nonumber\\
&&+\lambda_K \lambda_{J/\psi}\xi^{\alpha } \lambda_{Y}
\left(\zeta^{*\mu }p^{\prime\nu }-\zeta^{*\nu }p^{\prime\mu }\right) \frac{-iG_{Y J/\psi K}\varepsilon^{\rho\sigma\lambda\tau}p_\rho\xi^{*}_\sigma p^{\prime}_\lambda\zeta_\tau }{(M_{Y}^2-p^{\prime2})(M_{J/\psi}^2-p^2)(M_{K}^2-q^2)} \nonumber \\
&&+\cdots \, ,
\end{eqnarray}

\begin{eqnarray}\label{Hadron-CT-3}
\Pi^{\alpha\mu\nu}_2(p,q)&=& \lambda_\eta \lambda_{K^*}\xi^{\alpha } \lambda_{Z}
\varepsilon^{\mu\nu\mu^\prime\nu^\prime}\zeta^{*}_{\mu^\prime}p^{\prime}_{\nu^\prime} \frac{-G_{Z\eta K^*}\xi^{*} \cdot  \zeta }{(M_{Z}^2-p^{\prime2})(M_{\eta}^2-p^2)(M_{K^*}^2-q^2)}  \nonumber\\
&&+\lambda_\eta \lambda_{K^*}\xi^{\alpha } \lambda_{Y}
\left(\zeta^{*\mu }p^{\prime\nu }-\zeta^{*\nu }p^{\prime\mu }\right) \frac{-iG_{Y \eta K^*}\varepsilon^{\rho\sigma\lambda\tau}q_\rho\xi^{*}_\sigma p^{\prime}_\lambda\zeta_\tau }{(M_{Y}^2-p^{\prime2})(M_{\eta}^2-p^2)(M_{K^*}^2-q^2)} \nonumber \\
&&+\cdots \, ,
\end{eqnarray}
where $\lambda_K=\frac{f_{K}M_{K}^2}{m_u+m_s}$, $\lambda_\eta=\frac{f_{\eta_c}M_{\eta_c}^2}{2m_c}$, $\lambda_{J/\psi}=f_{J/\psi}M_{J/\psi}$, $\lambda_{K^*}=f_{K^*}M_{K^*}$, $p^\prime=p+q$, the decay constants of the mesons $h_c$, $J/\psi$, $K$, $\eta_c$, $K^*$  are defined by,
\begin{eqnarray}
\langle0|J_{\mu\nu}^{h_c}(0)|h_c(p)\rangle&=&f_{h_c} \varepsilon_{\mu\nu\alpha\beta}\xi^\alpha p^\beta \,\, , \nonumber \\
\langle0|J_{\mu\nu}^{h_c}(0)|J/\psi(p)\rangle&=&f_{J/\psi}^T \left(\xi_{\mu} p_\nu-\xi_{\nu} p_\mu\right) \,\, , \nonumber \\
\langle0|J_{\mu}^{J/\psi}(0)|J/\psi(p)\rangle&=&f_{J/\psi}M_{J/\psi}\,\xi_\mu \,\, , \nonumber \\
\langle0|J_{\mu}^{K^*}(0)|K^*(p)\rangle&=&f_{K^*}M_{K^*}\,\xi_\mu \,\, , \nonumber \\
\langle0|J_{5}^{K}(0)|K(p)\rangle&=&\frac{f_{K}M_{K}^2}{m_u+m_s} \,\, , \nonumber \\
\langle0|J_{5}^{\eta_c}(0)|\eta_c(p)\rangle&=&\frac{f_{\eta_c}M_{\eta_c}^2}{2m_c} \,\, ,
\end{eqnarray}
 the $\xi$  are polarization vectors of the $h_c$, $J/\psi$ and  $K^*$,  and the hadronic coupling constants are defined by
\begin{eqnarray}
\langle h_c(p)K(q)|Z_{cs}(p^{\prime})\rangle&=& G_{Zh K}\,\varepsilon^{\rho\sigma\lambda\tau}p_\rho\xi^{*}_\sigma p^{\prime}_\lambda\zeta_\tau \, , \nonumber\\
\langle J/\psi(p)K(q)|Y_{cs}(p^{\prime})\rangle&=& G_{Y J/\psi K}\,\varepsilon^{\rho\sigma\lambda\tau}p_\rho\xi^{*}_\sigma p^{\prime}_\lambda\zeta_\tau \, , \nonumber\\
\langle h_c(p)K(q)|Y_{cs}(p^{\prime})\rangle&=& -iG_{Yh K}\,\xi^{*} \cdot \zeta \, , \nonumber\\
\langle J/\psi(p)K(q)|Z_{cs}(p^{\prime})\rangle&=& -iG_{Z J/\psi K}\,\xi^{*} \cdot \zeta \, , \nonumber\\
\langle\eta_c(p)K^*(q)|Z_{cs}(p^{\prime})\rangle&=&-iG_{Z \eta K^*}\,\xi^{*} \cdot \zeta   \, .
\end{eqnarray}
The tensor structures in Eqs.\eqref{Hadron-CT-1}-\eqref{Hadron-CT-3} are enough complex, we have to project out the relevant  components with the suitable tensor operators,
\begin{eqnarray}
-\frac{2i}{9}\left(p^2q^2-(p\cdot q)^2 \right)\Pi_{h_cK}(p^{\prime2},p^2,q^2)&=&P_{A,p}^{\alpha\beta\eta\theta}P_{A,p^\prime}^{\mu\nu\phi\omega}\varepsilon_{\eta\theta\phi\omega}\Pi_{\alpha\beta\mu\nu}(p,q)\, , \nonumber \\
-6\left(p^2+q^2+2p\cdot q\right)\Pi_{J/\psi K}(p^{\prime2},p^2,q^2)&=&\varepsilon_{\mu\nu\alpha\sigma}p^{\prime \sigma}\Pi_1^{\alpha\mu\nu}(p,q)\, , \nonumber \\
-6\left(p^2+q^2+2p\cdot q\right)\Pi_{\eta_c K^*}(p^{\prime2},p^2,q^2)&=&\varepsilon_{\mu\nu\alpha\sigma}p^{\prime \sigma}\Pi_2^{\alpha\mu\nu}(p,q)\, ,
\end{eqnarray}
where
\begin{eqnarray}\label{H-Reprisen}
\Pi_{h_cK}(p^{\prime2},p^2,q^2)&=& \frac{G_{Zh K}\lambda_K f_{h} \lambda_{Z}}{(M_{Z}^2-p^{\prime2})(M_{h}^2-p^2)(M_{K}^2-q^2)}+\cdots \, , \nonumber\\
\Pi_{J/\psi K}(p^{\prime2},p^2,q^2)&=& \frac{G_{ZJ/\psi K}\lambda_K \lambda_{J/\psi} \lambda_{Z}}{(M_{Z}^2-p^{\prime2})(M_{J/\psi}^2-p^2)(M_{K}^2-q^2)}+\cdots \, ,\nonumber\\
\Pi_{\eta_c K^*}(p^{\prime2},p^2,q^2)&=& \frac{G_{Z\eta K^*}\lambda_{K^*} \lambda_{\eta} \lambda_{Z}}{(M_{Z}^2-p^{\prime2})(M_{\eta}^2-p^2)(M_{K^*}^2-q^2)}+\cdots \, ,
\end{eqnarray}
which correspond to the two-body strong decays $Z_{cs}\to h_c K$, $J/\psi K$, $\eta_c K^*$, respectively;
the other components in  Eqs.\eqref{Hadron-CT-1}-\eqref{Hadron-CT-3}  have no contributions or contaminations.
In Eq.\eqref{Hadron-CT-1}, there are four channels, $Z_{cs}\to h_c K$, $Y_{cs}\to h_c K$, $Z_{cs}\to J/\psi K$ and $Y_{cs}\to J/\psi K$, which correspond to four different tensor structures and therefore four different components,  we project out the
 channel $Z_{cs}\to h_c K$ explicitly.  In Eq.\eqref{Hadron-CT-2}, there are two channels $Z_{cs}\to J/\psi K$ and $Y_{cs}\to J/\psi K$,  which correspond to two different tensor structures and therefore two different components,  we project out the channel $Z_{cs}\to J/\psi K$ explicitly. In Eq.\eqref{Hadron-CT-3}, there are two channels $Z_{cs}\to \eta_cK^*$ and $Y_{cs}\to \eta_cK^*$, which correspond to two different tensor structures and therefore two different components, we project out the channel $Z_{cs}\to \eta_cK^*$ explicitly.
 The $\cdots$ in Eq.\eqref{H-Reprisen} stands for the neglected contributions from the higher resonances and continuum states.   According to the analysis in Refs.\cite{WZG-Zcs3985-decay,WZG-ZJX-Zc-Decay,WZG-Y4660-Decay,WZG-X4140-decay,WZG-X4274-decay,WZG-Z4600-decay}, we can introduce the parameters $C_{h_cK}$, $C_{J/\psi K}$ and $C_{\eta_cK^*}$ to parameterize the higher resonance and continuum states involving  the $Z_{cs}$ channel,
\begin{eqnarray}\label{H-ChcK-etal}
\Pi_{h_cK}(p^{\prime2},p^2,q^2)&=& \frac{G_{Zh K}\lambda_K f_{h} \lambda_{Z}}{(M_{Z}^2-p^{\prime2})(M_{h}^2-p^2)(M_{K}^2-q^2)}+\frac{C_{h_c K}}{(M_{h}^2-p^2)(M_{K}^2-q^2)} \, , \nonumber\\
\Pi_{J/\psi K}(p^{\prime2},p^2,q^2)&=& \frac{G_{ZJ/\psi K}\lambda_K \lambda_{J/\psi} \lambda_{Z}}{(M_{Z}^2-p^{\prime2})(M_{J/\psi}^2-p^2)(M_{K}^2-q^2)}
+\frac{C_{J/\psi K}}{(M_{J/\psi}^2-p^2)(M_{K}^2-q^2)} \, ,\nonumber\\
\Pi_{\eta_c K^*}(p^{\prime2},p^2,q^2)&=& \frac{G_{Z\eta K^*}\lambda_{K^*} \lambda_{\eta} \lambda_{Z}}{(M_{Z}^2-p^{\prime2})(M_{\eta}^2-p^2)(M_{K^*}^2-q^2)}+\frac{C_{\eta_c K^*}}{(M_{\eta}^2-p^2)(M_{K^*}^2-q^2)} \, .
\end{eqnarray}

On the other hand, we perform Fierz re-arrangement both in the color and Dirac-spinor  spaces to obtain the  result,
\begin{eqnarray}
2\sqrt{2} J_{u\bar{s}}^{\mu\nu} &=&i\bar{s}u\, \bar{c}\sigma^{\mu\nu}c +i\bar{s}\sigma^{\mu\nu}u \,\bar{c}c+i\bar{s}c\, \bar{c}\sigma^{\mu\nu}u +i\bar{s}\sigma^{\mu\nu}c \,\bar{c}u -\frac{i}{2}\varepsilon^{\mu\nu\alpha\beta}\bar{c}\sigma_{\alpha\beta}c\,\bar{s}i\gamma_5u\nonumber\\
 &&-\bar{c}i\gamma_5 c\,\bar{s}\sigma^{\mu\nu}\gamma_5u -\bar{c}\sigma^{\mu\nu}\gamma_5u\,\bar{s}i\gamma_5c-\bar{s}i\gamma_5 c\,\bar{c}\sigma^{\mu\nu}\gamma_5u+i\varepsilon^{\mu\nu\alpha\beta}\bar{c}\gamma^\alpha\gamma_5c\, \bar{s}\gamma^\beta u\nonumber\\
 &&-i\varepsilon^{\mu\nu\alpha\beta}\bar{c}\gamma^\alpha c\, \bar{s}\gamma^\beta \gamma_5u+i\varepsilon^{\mu\nu\alpha\beta}\bar{c}\gamma^\alpha\gamma_5u\, \bar{s}\gamma^\beta c-i\varepsilon^{\mu\nu\alpha\beta}\bar{c}\gamma^\alpha u\, \bar{s}\gamma^\beta \gamma_5c \, ,
\end{eqnarray}
the component $\frac{i}{2}\varepsilon^{\mu\nu\alpha\beta}\bar{c}\sigma_{\alpha\beta}c\,\bar{s}i\gamma_5u$ leads  to the correlation function,
\begin{eqnarray}
\widetilde{\Pi}_{\alpha\beta\mu\nu}(p,q)&=&\frac{i^2\varepsilon_{\mu\nu\lambda\tau}}{4\sqrt{2}}\int d^4xd^4y \, e^{ipx}e^{iqy}\, \langle 0|T\left\{J_{\alpha\beta}^{h_c}(x)J_5^{K}(y)\, \bar{c}(0)\sigma^{\lambda\tau}c(0)\,\bar{u}(0)i\gamma_5s(0)\right\}|0\rangle\, ,\nonumber\\
&\to & \kappa \frac{i^2\varepsilon_{\mu\nu\lambda\tau}}{4\sqrt{2}}\int d^4x \, e^{ipx}\, \langle 0|T\left\{J_{\alpha\beta}^{h_c}(x)\, \bar{c}(0)\sigma^{\lambda\tau}c(0)\right\}|0\rangle \nonumber \\
&&\int d^4y \, e^{iqy}\, \langle 0|T\left\{J_5^{K}(y)\, \bar{u}(0)i\gamma_5s(0)\right\}|0\rangle\, ,
\end{eqnarray}
we introduce a parameter $\kappa$ to represent the possible factorizable  contributions at the hadron side, as we choose the local currents, and the conventional mesons  and tetraquark states have average  spatial sizes of the same order,  the $J_{u\bar{s}}^{\mu\nu}(0)$ couples  potentially to the tetraquark state rather than to the two-meson scattering states, therefore $\kappa \ll 1$ \cite{WZG-X4140-X4685}, however, such a term makes a contribution to the component  $\Pi_{h_cK}(p^{\prime2},p^2,q^2)$,
\begin{eqnarray}
\frac{\widetilde{C}_{h_c K}}{(M_{h}^2-p^2)(M_{K}^2-q^2)}\, ,
\end{eqnarray}
where the coefficient $\widetilde{C}_{h_c K}$ can be absorbed into the coefficient $C_{h_cK}$. We can see confidently that the parameter $C_{h_cK}$ is necessary, the parameters $C_{J/\psi K}$ and $C_{\eta_cK^*}$ are implied in the same way.

 We accomplish  the operator product expansion up to the vacuum condensates of dimension 5 and neglect the tiny gluon condensate contributions \cite{WZG-Zcs3985-decay,WZG-ZJX-Zc-Decay,WZG-Y4660-Decay,WZG-X4140-decay,WZG-X4274-decay,WZG-Z4600-decay}, then obtain the QCD spectral densities $\rho_{QCD}(p^{\prime2},s,u)$  through double dispersion relation,
\begin{eqnarray}
\Pi_{QCD}(p^{\prime2},p^2,q^2)&=& \int_{\Delta_s^2}^\infty ds \int_{\Delta_u^2}^\infty du \frac{\rho_{QCD}(p^{\prime2},s,u)}{(s-p^2)(u-q^2)}\, ,
\end{eqnarray}
where   the $\Delta_s^2$ and $\Delta_u^2$  are the thresholds.
At the hadron side, we obtain the hadronic   spectral densities $\rho_H(s^\prime,s,u)$ through triple  dispersion relation,
\begin{eqnarray}
\Pi_{H}(p^{\prime2},p^2,q^2)&=&\int_{\Delta_s^{\prime2}}^\infty ds^{\prime} \int_{\Delta_s^2}^\infty ds \int_{\Delta_u^2}^\infty du \frac{\rho_{H}(s^\prime,s,u)}{(s^\prime-p^{\prime2})(s-p^2)(u-q^2)}\, ,
\end{eqnarray}
according to Eq.\eqref{H-Reprisen}, where the $\Delta_{s}^{\prime2}$ are the thresholds.
We match the hadron side with the QCD side  bellow the continuum thresholds  to acquire   rigorous quark-hadron  duality  \cite{WZG-ZJX-Zc-Decay,WZG-Y4660-Decay},
 \begin{eqnarray}
  \int_{\Delta_s^2}^{s_{0}}ds \int_{\Delta_u^2}^{u_0}du  \frac{\rho_{QCD}(p^{\prime2},s,u)}{(s-p^2)(u-q^2)}&=& \int_{\Delta_s^2}^{s_0}ds \int_{\Delta_u^2}^{u_0}du  \left[ \int_{\Delta_{s}^{\prime2}}^{\infty}ds^\prime  \frac{\rho_H(s^\prime,s,u)}{(s^\prime-p^{\prime2})(s-p^2)(u-q^2)} \right]\, ,
\end{eqnarray}
where  the $s_0$ and $u_0$ are the continuum thresholds,  we accomplish the integral over $ds^\prime$ firstly, and introduce some unknown parameters, such as the $C_{h_cK}$, $C_{J/\psi K}$ and $C_{\eta_cK^*}$, to parameterize the contributions involving the higher resonances and continuum states in the $s^\prime$ channel.

We set $p^{\prime2}=p^2$ in the correlation functions $\Pi(p^{\prime 2},p^2,q^2)$, and accomplish the double Borel transform in regard  to the variables $P^2=-p^2$ and $Q^2=-q^2$ respectively, then set the Borel parameters  $T_1^2=T_2^2=T^2$  to obtain  three QCD sum rules,
\begin{eqnarray} \label{ZhcK-SR}
&&\frac{\lambda_{Zh K}G_{Zh K}}{M_{Z}^2-M_{h}^2} \left[ \exp\left(-\frac{M_{h}^2}{T^2} \right)-\exp\left(-\frac{M_{Z}^2}{T^2} \right)\right]\exp\left(-\frac{M_{K}^2}{T^2} \right)+C_{h_c K} \exp\left(-\frac{M_{h}^2+M_{K}^2}{T^2}  \right) \nonumber\\
&&=\frac{1}{64\sqrt{2}\pi^4}\int_{4m_c^2}^{s^0_{h}} ds \int_{0}^{s^0_{K}} du  \sqrt{1-\frac{4m_c^2}{s}} \left(1-\frac{4m_c^2}{s}\right)\exp\left(-\frac{s+u}{T^2}\right)\nonumber\\
&&+\frac{m_s\left[2\langle \bar{q}q\rangle-\langle\bar{s}s\rangle\right]}{48\sqrt{2}\pi^2 T^2}\int_{4m_c^2}^{s^0_{h}} ds \sqrt{1-\frac{4m_c^2}{s}} \left(1-\frac{4m_c^2}{s}\right)\exp\left(-\frac{s}{T^2}\right)\nonumber\\
&&+\frac{m_s \langle \bar{q}  G q \rangle}{96\sqrt{2} \pi^2 T^2} \int_{4m_c^2}^{s^0_{h}} ds \frac{1}{\sqrt{s \left(s-4m_c^2\right)}} \left(1-\frac{2 m_c^2}{s}\right) \exp\left(-\frac{s}{T^2}\right) \nonumber\\
&&+\frac{m_s \langle \bar{q}  G q \rangle}{96\sqrt{2} \pi^2 T^2} \int_{4m_c^2}^{s^0_{h}} ds \sqrt{1-\frac{4m_c^2}{s}}  \frac{1}{s} \exp\left(-\frac{s}{T^2}\right) \nonumber\\
&&+\frac{m_s \langle \bar{q}  G q \rangle}{64 \sqrt{2} \pi^2 T^4} \int_{4m_c^2}^{s^0_{h}} ds \sqrt{1-\frac{4m_c^2}{s}} \left(1-\frac{4m_c^2}{s}\right) \exp\left(-\frac{s}{T^2}\right) \, ,
\end{eqnarray}

\begin{eqnarray} \label{ZJpsiK-SR}
&&\frac{\lambda_{ZJ/\psi K}G_{ZJ/\psi K}}{M_{Z}^2-M_{J/\psi}^2} \left[ \exp\left(-\frac{M_{J/\psi}^2}{T^2} \right)-\exp\left(-\frac{M_{Z}^2}{T^2} \right)\right]\exp\left(-\frac{M_{K}^2}{T^2} \right)+C_{J/\psi K} \exp\left(-\frac{M_{J/\psi}^2+M_{K}^2}{T^2}  \right) \nonumber\\
&&=\frac{3}{128\sqrt{2}\pi^4}\int_{4m_c^2}^{s^0_{J/\psi}} ds \int_{0}^{s^0_{K}} du  \sqrt{1-\frac{4m_c^2}{s}} \left[2um_c+m_s\left(s+2m_c^2\right)\left(\frac{2}{3}-\frac{u}{9s}\right) \right]\exp\left(-\frac{s+u}{T^2}\right)\nonumber\\
&&-\frac{\langle \bar{q}q\rangle+\langle\bar{s}s\rangle}{24\sqrt{2}\pi^2} \int_{4m_c^2}^{s^0_{J/\psi}} ds \sqrt{1-\frac{4m_c^2}{s}} \left(s+2m_c^2\right) \exp\left(-\frac{s}{T^2}\right) \nonumber\\
&&+\frac{m_s m_c\left[\langle\bar{s}s\rangle-2\langle \bar{q}q\rangle\right]}{16\sqrt{2}\pi^2} \int_{4m_c^2}^{s^0_{J/\psi}} ds \sqrt{1-\frac{4m_c^2}{s}} \exp\left(-\frac{s}{T^2}\right) \nonumber\\
&&+\frac{\langle \bar{q}  G q \rangle+\langle \bar{s}  G s\rangle}{576\sqrt{2} \pi^2} \int_{4m_c^2}^{s^0_{J/\psi}} ds \frac{s+8 m_c^2}{\sqrt{s \left(s-4m_c^2\right)}}  \exp\left(-\frac{s}{T^2}\right) -\frac{\langle \bar{q}  G q \rangle+\langle \bar{s}  G s\rangle}{576\sqrt{2} \pi^2} \int_{4m_c^2}^{s^0_{J/\psi}} ds \sqrt{1-\frac{4m_c^2}{s}} \exp\left(-\frac{s}{T^2}\right)  \nonumber\\
&&+\frac{m_s m_c\langle \bar{q}  G q \rangle}{192\sqrt{2} \pi^2} \int_{4m_c^2}^{s^0_{J/\psi}} ds \frac{1}{\sqrt{s \left(s-4m_c^2\right)}} \exp\left(-\frac{s}{T^2}\right) -\frac{m_s m_c\langle \bar{q}  G q \rangle}{192\sqrt{2} \pi^2} \int_{4m_c^2}^{s^0_{J/\psi}} ds \sqrt{1-\frac{4m_c^2}{s}} \frac{1}{s} \exp\left(-\frac{s}{T^2}\right) \nonumber\\
&& -\frac{m_s m_c\langle \bar{q}  G q \rangle}{16\sqrt{2} \pi^2 T^2} \int_{4m_c^2}^{s^0_{J/\psi}} ds \sqrt{1-\frac{4m_c^2}{s}} \exp\left(-\frac{s}{T^2}\right) \, ,
\end{eqnarray}

\begin{eqnarray}\label{ZetacKV-SR}
&& \frac{\lambda_{Z \eta K^*}G_{Z \eta K^*}}{M_{Z}^2-M_{\eta}^2}\left[ \exp\left(-\frac{M_{\eta}^2}{T^2} \right)-\exp\left(-\frac{M_{Z}^2}{T^2} \right)\right]\exp\left(-\frac{M_{K^*}^2}{T^2} \right)+C_{\eta_c K^*} \exp\left(-\frac{M_{\eta}^2+M_{K^*}^2}{T^2}  \right) \nonumber\\
&&=\frac{3}{128\sqrt{2}\pi^4}\int_{4m_c^2}^{s^0_{\eta_c}} ds \int_{0}^{s^0_{K^*}} du \sqrt{1-\frac{4m_c^2}{s}} \left(\frac{10um_c}{9}+m_s s \right)\exp\left(-\frac{s+u}{T^2}\right)\nonumber\\
&&-\frac{\langle \bar{q}q\rangle+\langle\bar{s}s\rangle}{16\sqrt{2}\pi^2} \int_{4m_c^2}^{s^0_{\eta_c}} ds \sqrt{1-\frac{4m_c^2}{s}}\, s\, \exp\left(-\frac{s}{T^2}\right) \nonumber\\
&&+\frac{m_s m_c\left[\langle\bar{s}s\rangle-6\langle \bar{q}q\rangle\right]}{48\sqrt{2}\pi^2} \int_{4m_c^2}^{s^0_{\eta_c}} ds \sqrt{1-\frac{4m_c^2}{s}} \exp\left(-\frac{s}{T^2}\right)\nonumber\\
&&+\frac{\langle \bar{q}  G q \rangle+\langle \bar{s}  G s\rangle}{576\sqrt{2} \pi^2} \int_{4m_c^2}^{s^0_{\eta_c}} ds \frac{s+2m_c^2}{\sqrt{s \left(s-4m_c^2\right)}}   \exp\left(-\frac{s}{T^2}\right) \nonumber\\
&& -\frac{\langle \bar{q}  G q \rangle+\langle \bar{s}  G s\rangle}{576\sqrt{2} \pi^2} \int_{4m_c^2}^{s^0_{\eta_c}} ds \sqrt{1-\frac{4m_c^2}{s}} \left(1-\frac{12s}{T^2}\right) \exp\left(-\frac{s}{T^2}\right) \nonumber\\
&&+\frac{m_s m_c\langle \bar{q}  G q \rangle}{96\sqrt{2} \pi^2} \int_{4m_c^2}^{s^0_{\eta_c}} ds \frac{1}{\sqrt{s \left(s-4m_c^2\right)}} \exp\left(-\frac{s}{T^2}\right) \nonumber\\
&&+\frac{m_s m_c\langle \bar{s}  G s \rangle}{288\sqrt{2} \pi^2 T^2} \int_{4m_c^2}^{s^0_{\eta_c}} ds \sqrt{1-\frac{4m_c^2}{s}} \exp\left(-\frac{s}{T^2}\right) \, ,
\end{eqnarray}
where $\langle \bar{q}Gq\rangle=\langle \bar{q}g_s \sigma Gq\rangle$, $\langle \bar{s}Gs\rangle=\langle \bar{s}g_s \sigma Gs\rangle$,
$\lambda_{Zh K}=\lambda_{K}f_{h}\lambda_{Z}$, $\lambda_{ZJ/\psi K}=\lambda_{K}\lambda_{J/\psi}\lambda_{Z}$, $\lambda_{Z\eta K^*}=\lambda_{K^*}\lambda_{\eta}\lambda_{Z}$. We neglect  the dependencies of the parameters  $C_{h_c K}$, $C_{J/\psi K}$ and $C_{\eta_c K^*}$ on the Lorentz invariants $p^{\prime2}$, $p^2$, $q^2$,  take them  as free parameters, and search for the best values  to
delete  the contaminations from the high resonances and continuum states to acquire  stable QCD sum rules. The corresponding hadronic coupling constants for the $Z_c(4020/4025)$ state  can be obtained with the simple replacement $s \to d$ and are treated in the same way.

At the QCD side, we choose the flavor numbers $n_f=4$ and set the energy scale to be $\mu=1.3\,\rm{GeV}$, just like in previous work on the decays of the  $Z_{cs}(3985/4000)$ \cite{WZG-Zcs3985-decay}. At the hadron side, we take the parameters  as $M_{K}=0.4937\,\rm{GeV}$, $M_{\pi}=0.13957\,\rm{GeV}$, $M_{K^*}=0.8917\,\rm{GeV}$, $M_{\rho}=0.77526\,\rm{GeV}$,
$M_{J/\psi}=3.0969\,\rm{GeV}$, $M_{\eta_c}=2.9834\,\rm{GeV}$, $M_{h_c}=3.525\,\rm{GeV}$ \cite{PDG},  $f_{K}=0.156\,\rm{GeV}$, $f_{\pi}=0.130\,\rm{GeV}$  \cite{PDG}, $f_{K^*}=0.220\,\rm{GeV}$, $f_{\rho}=0.215\,\rm{GeV}$, $\sqrt{s^0_{K}}=1.0\,\rm{GeV}$, $\sqrt{s^0_{\pi}}=0.85\,\rm{GeV}$, $\sqrt{s^0_{K^*}}=1.3\,\rm{GeV}$, $\sqrt{s^0_{\rho}}=1.2\,\rm{GeV}$ \cite{PBall-decay-Kv},
  $f_{h_c}=0.235\,\rm{GeV}$, $f_{J/\psi}=0.418 \,\rm{GeV}$, $f_{\eta_c}=0.387 \,\rm{GeV}$  \cite{Becirevic}, $\sqrt{s^0_{h_c}}=4.05\,\rm{GeV}$, $\sqrt{s^0_{J/\psi}}=3.6\,\rm{GeV}$, $\sqrt{s^0_{\eta_c}}=3.5\,\rm{GeV}$ \cite{PDG},  $\frac{f_{K}M^2_{K}}{m_u+m_s}=-\frac{\langle \bar{q}q\rangle+\langle \bar{s}s\rangle}{f_{K}(1-\delta_K)}$, $\frac{f_{\pi}M^2_{\pi}}{m_u+m_d}=-\frac{2\langle \bar{q}q\rangle}{f_{\pi}}$ from the Gell-Mann-Oakes-Renner relation, $\delta_K=0.50$ \cite{GMOR-fK}.

\begin{figure}
\centering
\includegraphics[totalheight=6cm,width=7cm]{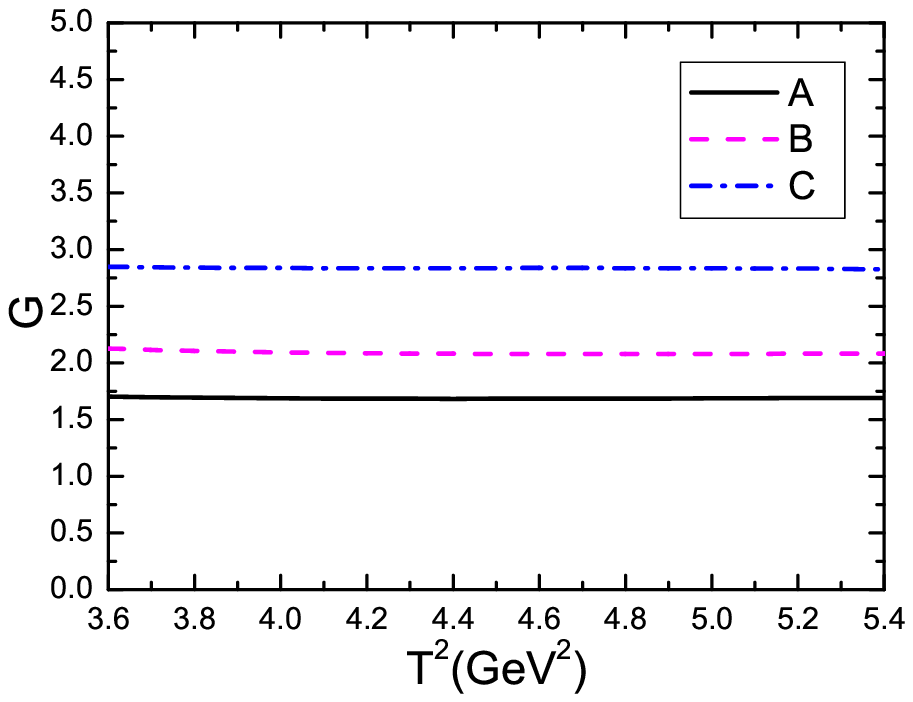}
\includegraphics[totalheight=6cm,width=7cm]{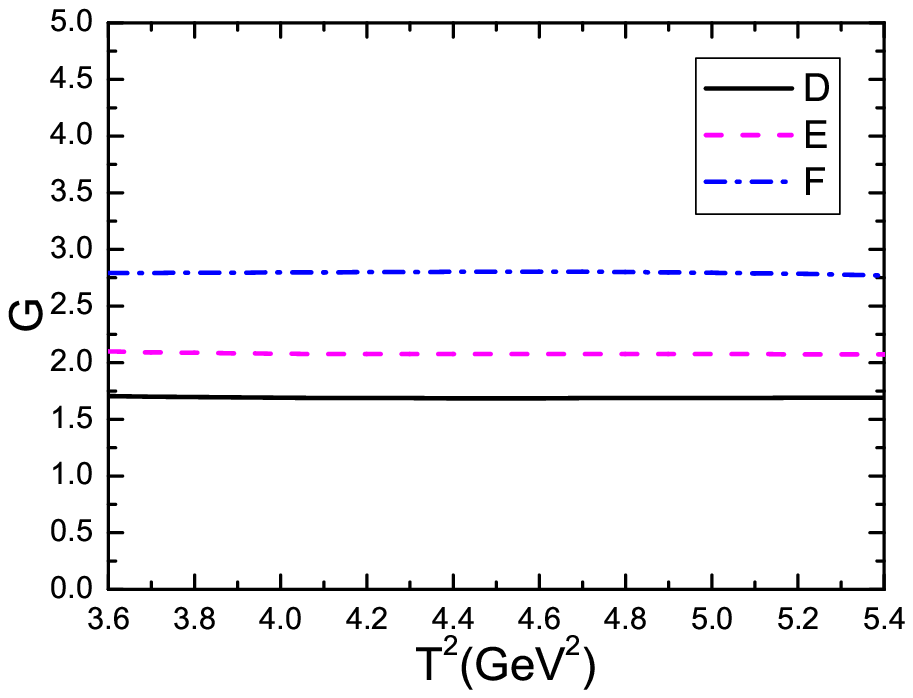}
  \caption{ The hadronic coupling constants with variations of the Borel parameters $T^2$, where the $A$, $B$, $C$, $D$, $E$ and $F$ correspond  to the $G_{Z_{cs} h_c K}$, $G_{Z_{cs}J/\psi K}$, $G_{Z_{cs}\eta_c K^*}$, $G_{Z_{c} h_c \pi}$, $G_{Z_{c}J/\psi \pi}$ and $G_{Z_{c}\eta_c \rho}$, respectively.    }\label{hadron-coupling}
\end{figure}

In calculations, we fit the unknown parameters to be
$C_{h_c K}=0.000064+0.000014\times T^2\,\rm{GeV}^4$,
$C_{h_c \pi}=0.00006+0.000010\times T^2\,\rm{GeV}^4$,
$C_{J/\psi K}=0.00335+0.000096\times T^2\,\rm{GeV}^7$,
$C_{J/\psi \pi}=0.00305+0.000096\times T^2\,\rm{GeV}^7$,
 $C_{\eta_c K^*}=0.00368+0.00012\times T^2\,\rm{GeV}^7$ and
 $C_{\eta_c \rho}=0.00302+0.00012\times T^2\,\rm{GeV}^7$
  to acquire the flat Borel platforms having the interval  $T^2_{max}-T^2_{min}=1\,\rm{GeV}^2$, where the max and min represent the maximum and minimum values, respectively.
The Borel windows  are $T^2_{h_c K}=(4.0-5.0)\,\rm{GeV}^2$, $T^2_{h_c \pi}=(4.0-5.0)\,\rm{GeV}^2$,
$T^2_{J/\psi K}=(4.3-5.3)\,\rm{GeV}^2$, $T^2_{J/\psi \pi}=(4.1-5.1)\,\rm{GeV}^2$,
$T^2_{\eta_c K^*}=(3.9-4.9)\,\rm{GeV}^2$ and $T^2_{\eta_c \rho}=(3.9-4.9)\,\rm{GeV}^2$, where we add the subscripts $h_cK$, $h_c\pi$ $\cdots$ to denote the corresponding decay channels. In the Borel windows, we require that the uncertainties  $\delta G$ originate from the Borel parameters $T^2$ are less than or about $0.01\, (\rm GeV)$, such a strict and powerful constraint plays a decisive role and works well, just like in our previous works \cite{WZG-Zcs3985-decay,WZG-ZJX-Zc-Decay,WZG-Y4660-Decay,WZG-X4140-decay,WZG-X4274-decay,WZG-Z4600-decay}. In Fig.\ref{hadron-coupling}, we plot the hadronic coupling constants $G_{Z_{cs} h_c K}$, $G_{Z_{cs}J/\psi K}$, $G_{Z_{cs}\eta_c K^*}$, $G_{Z_{c} h_c \pi}$, $G_{Z_{c}J/\psi \pi}$ and $G_{Z_{c}\eta_c \rho}$ with variations of the Borel parameters, where we can see explicitly that there appear very flat platforms indeed, it is reliable to extract the hadronic coupling constants.

If we take  the symbol   $\xi$ to represent the input parameters at the QCD side, then for example,  the uncertainties   $\bar{\xi} \to \bar{\xi} +\delta \xi$ result in the uncertainties $\bar{f}_{J/\psi}\bar{f}_{K}\bar{\lambda}_{Z}\bar{G}_{ZJ/\psi K} \to \bar{f}_{J/\psi}\bar{f}_{K}\bar{\lambda}_{Z}\bar{G}_{ZJ/\psi K}+\delta\,f_{J/\psi}f_{K}\lambda_{Z}G_{ZJ/\psi K}$, $\bar{C}_{J/\psi K} \to \bar{C}_{J/\psi K}+\delta C_{J/\psi K}$,
where
\begin{eqnarray}\label{Uncertainty-4}
\delta\,f_{J/\psi}f_{K}\lambda_{Z}G_{ZJ/\psi K} &=&\bar{f}_{J/\psi}\bar{f}_{K}\bar{\lambda}_{Z}\bar{G}_{ZJ/\psi K}\left( \frac{\delta f_{J/\psi}}{\bar{f}_{J/\psi}} +\frac{\delta f_{K}}{\bar{f}_{K}}+\frac{\delta \lambda_{Z}}{\bar{\lambda}_{Z}}+\frac{\delta G_{ZJ/\psi K}}{\bar{G}_{ZJ/\psi K}}\right)\, ,
\end{eqnarray}
 we add the index \,$\bar{}$\, on all the variables  to denote the central values.
In the case of the uncertainty $\delta C_{J/\psi K}$ is small enough to  be  ignored, error analysis is easy to preform by setting  $\frac{\delta f_{J/\psi}}{\bar{f}_{J/\psi}} =\frac{\delta f_{K}}{\bar{f}_{K}}=\frac{\delta \lambda_{Z}}{\bar{\lambda}_{Z}}=\frac{\delta G_{ZJ/\psi K}}{\bar{G}_{ZJ/\psi K}}
$ approximately, on the other hand, if the uncertainty $\delta C_{J/\psi K}$ is considerable, we have to take it into account for every uncertainty $\delta \xi$. We have to adjust the $\delta C_{J/\psi K}$ with fine tuning with the help of  trial and error according to the variation $\delta \xi$ to acquire enough  flat platforms  in the same region, just like in the case of the central values $\bar{\xi}$ and $\bar{C}_{J/\psi K}$.  It is a difficult work to perform error analysis. We usually set $\frac{\delta f_{J/\psi}}{\bar{f}_{J/\psi}} =\frac{\delta f_{K}}{\bar{f}_{K}}=\frac{\delta \lambda_{Z}}{\bar{\lambda}_{Z}}=0$ to estimate the uncertainty $\delta G_{ZJ/\psi K}$, in fact, no one has ever proved that such an approximation is proper.

Now let us obtain the hadronic coupling constants routinely according to above error analysis,
\begin{eqnarray} \label{HCC-values}
G_{Z_{cs} h_c K} &=&1.68 \pm 0.10\, , \nonumber\\
G_{Z_{cs}J/\psi K} &=&2.08\pm 0.08\,\rm{GeV}\, , \nonumber\\
G_{Z_{cs}\eta_c K^*} &=&2.84\pm 0.09\,\rm{GeV}\, , \nonumber\\
G_{Z_{c} h_c \pi} &=&1.69\pm 0.09 \, , \nonumber\\
G_{Z_{c}J/\psi \pi} &=&2.08\pm 0.08\,\rm{GeV}\, , \nonumber\\
G_{Z_{c}\eta_c \rho} &=&2.80\pm 0.09\,\rm{GeV}\, ,
\end{eqnarray}
by setting
\begin{eqnarray}\label{Uncertainty-5}
\delta\,f_{J/\psi}f_{K}\lambda_{Z}G_{ZJ/\psi K} &=&\bar{f}_{J/\psi}\bar{f}_{K}\bar{\lambda}_{Z}\bar{G}_{ZJ/\psi K}\frac{4\delta G_{ZJ/\psi K}}{\bar{G}_{ZJ/\psi K}}\, ,
\end{eqnarray}
et al.
If we set
\begin{eqnarray}\label{Uncertainty-6}
\delta\,f_{J/\psi}f_{K}\lambda_{Z}G_{ZJ/\psi K} &=&\bar{f}_{J/\psi}\bar{f}_{K}\bar{\lambda}_{Z}\bar{G}_{ZJ/\psi K}\frac{\delta G_{ZJ/\psi K}}{\bar{G}_{ZJ/\psi K}}\, ,
\end{eqnarray}
the uncertainty $\delta G_{ZJ/\psi K}$ will be four times as large as that given in Eq.\eqref{HCC-values}. Other uncertainties can be understood in the same way. According to Eq.\eqref{HCC-values},
the $SU(3)$ breaking effects  in the hadronic coupling constants are rather small.

Then it is easy to obtain the partial decay widths by taking the relevant masses  from the Particle Data Group \cite{PDG},
\begin{eqnarray} \label{Partial withs}
\Gamma\left(Z_{cs}\to h_c K\right)&=&1.83\pm0.22\,\rm{MeV}\, , \nonumber\\
\Gamma\left(Z_{cs}\to J/\psi K\right) &=&8.05\pm0.62\,\rm{MeV}\, , \nonumber\\
\Gamma\left(Z_{cs}\to \eta_c K^*\right) &=&12.83\pm0.81\,\rm{MeV}\, , \nonumber\\
\Gamma\left(Z_{c} \to h_c \pi\right) &=&6.86\pm0.73\,\rm{MeV}\, , \nonumber\\
\Gamma\left(Z_{c}\to J/\psi \pi\right) &=&8.82\pm0.68\,\rm{MeV}\, , \nonumber\\
\Gamma\left(Z_{c}\to \eta_c \rho\right) &=&13.89\pm0.89\,\rm{MeV}\, ,
\end{eqnarray}
and the total widths,
\begin{eqnarray} \label{Total widths}
\Gamma_{Z_{cs}} &=& 22.71\pm1.65\, ({\rm or}\, \pm 6.60)\,\rm{MeV}\, ,\nonumber\\
\Gamma_{Z_{c}} &=&29.57\pm2.30\, ({\rm or}\, \pm 9.20)\,\,\rm{MeV}\, ,
\end{eqnarray}
the values in the brackets are obtained from Eq.\eqref{Uncertainty-6}.
The prediction $\Gamma_{Z_{c}} =29.57\pm2.30\, ({\rm or}\, \pm 9.20)\,\,\rm{MeV}$ is compatible with the upper  bound of the experimental data  $\Gamma=(24.8\pm5.6\pm7.7)\,\rm{MeV}$ \cite{BES1308},  $(23.0\pm 6.0\pm 1.0)\,\rm{MeV}$ \cite{BES1507}, $(7.9\pm 2.7\pm 2.6)\,\rm{MeV}$ \cite{BES1309} from the BESIII collaboration, and also supports assigning the $Z_c(4020/4025)$ to be the $A\bar{A}$-type hidden-charm tetraquark states with the $J^{PC}=1^{+-}$. In the present work, we have neglected the decays $Z_c(4020/4025)\to D^*\bar{D}^*$ and $Z_{cs}\to D^*\bar{D}_s^*$, $D_s^*\bar{D}^*$, as the $Z_c$ and $Z_{cs}$ states lie near the corresponding two-meson thresholds, the available phase-spaces are very small,   and even lead to the possible assignments of molecular states \cite{Molecule-1,Molecule-2,Molecule-3,Molecule-4,Molecule-5,Molecule-6,Molecule-7,ZhangJR3900}. The most favorable channels are $Z_{cs}\to \eta_c K^*$ and $Z_{c}\to \eta_c \rho$, at the present time, even for the $Z_c(4020/4025)$, the decay $Z_c(4020/4025) \to \eta_c\rho$ has not been observed yet, the observation of this channel can lead to more robust assignment and shed light on the nature of the $Z_c$ states. We can search for the $Z_{cs}$ state in the invariant mass spectrum of the $h_c K$, $J/\psi K$, $\eta_c K^*$, $D^*\bar{D}_s^*$, $D_s^*\bar{D}^*$ in the future.

In the picture of diquark-antidiquark type tetraquark states, the $Z_c(3900)$ and $Z_{cs}(3985)$ can be  assigned tentatively as the $S\bar{A}-A\bar{S}$ type hidden-charm tetraquark states, the hadronic coupling constants have the relations $|G_{ZD^*\bar{D}/ZD\bar{D}^*}|\ll |G_{ZJ/\psi\pi/Z\eta_c\rho}|$ and $|G_{ZD^*\bar{D}_s/ZD\bar{D}_s^*}|\ll |G_{ZJ/\psi K/Z\eta_c K^*}|$, furthermore, the allowed phase-spaces in the decays to the open-charm meson pairs are much smaller than that to the meson pairs involving charmonium, the contributions of the decays to the open-charm meson pairs to the total decay widths can be ignored \cite{WZG-Zcs3985-decay,WZG-ZJX-Zc-Decay}. We expect that the conclusion survives in the present work for the $Z_c(4020/4025)$ and $Z_{cs}(4110)$ states, and make a crude estimation of the partial decay widths $\Gamma\left(Z_{c}\to D^* \bar{D}/D \bar{D}^*\right)< 1\,\rm{MeV}$ and $\Gamma\left(Z_{cs}\to D^* \bar{D}_s/D \bar{D}_s^*\right)<1\,\rm{MeV}$ based on the relations between  the hadronic coupling constants obtained in Refs.\cite{WZG-Zcs3985-decay,WZG-ZJX-Zc-Decay}, the contributions to the total widths from the decays to the final states $D^* \bar{D}/D \bar{D}^*$ and $D^* \bar{D}_s/D \bar{D}_s^*$ are also ignored.

\section{Conclusion}
In this article, we tentatively assign the $Z_c(4020/4025)$ as the $A\bar{A}$-type hidden-charm tetraquark state with the $J^{PC}=1^{+-}$, construct the $A\bar{A}$-type tensor currents to investigate the tetraquark states without strange, with strange and with hidden-strange together via the QCD sum rules. We take account of the contributions of the vacuum condensates up to dimension-10  in the operator product expansion,  then we resort to the modified energy scale  formula $\mu=\sqrt{M^2_{X/Y/Z}-(2{\mathbb{M}}_c)^2}-k{\mathcal{M}}_s$ to account for the $SU(3)$ mass-breaking effects to choose the suitable energy scales of the QCD spectral densities, and obtain the tetraquark masses in a self-consistent way. We introduce three-point correlation functions to investigate the hadronic  coupling constants in the two-body strong decays of the tetraquark states without strange and with strange together via the QCD sum rules  based on rigorous quark-hadron duality, it is the unique feature of our works.     The numerical results indicate that the $SU(3)$ breaking effects  in the hadronic coupling constants are rather small. Then we obtain the partial decay widths and total widths for the $Z_c$ and $Z_{cs}$ states, the total width  $\Gamma_{Z_c}$ is compatible with that of the $Z_c(4020/4025)$ and   also supports assigning the $Z_c(4020/4025)$ to be the $J^{PC}=1^{+-}$ $A\bar{A}$-type tetraquark state, more experimental data are still needed to reach more robust  assignment,  as the $Z_c(4020/4025)$ has not been observed in the $J/\psi \pi$ and $\eta_c\rho$ channels yet.
We can search for the strange cousin $Z_{cs}$ in the $D^*\bar{D}_s^*$, $D_s^*\bar{D}^*$, $h_cK$, $J/\psi K$ and $\eta_c K^*$ invariant mass spectrum in the future, its observation would shed light on the nature of the $Z_c$ states.

\section*{Acknowledgements}
This  work is supported by National Natural Science Foundation, Grant Number  12175068.

\end{document}